\documentclass[aps,prd,12point,twocolumn,nofootinbib,showpacs,superscriptaddress]{revtex4-2}
\def\theequation{\arabic{section}.\arabic{equation}}
\pdfoutput=1
\usepackage{psfrag}
\usepackage{subfigure}
\usepackage{cancel}
\usepackage[dvipsnames]{xcolor}
\usepackage{color}
\usepackage{mathrsfs}
\usepackage{graphicx}
\usepackage{amssymb,bm}

\usepackage{amsmath,amsthm}
\usepackage{epstopdf}
\usepackage[bookmarks]{hyperref}
\usepackage{enumerate}
\usepackage{longtable}

\hypersetup{
    colorlinks=true,
    linkcolor=blue,
    filecolor=magenta,
    urlcolor=cyan,
    citecolor=magenta
}
\newcommand{\be}{\begin{equation}}
\newcommand{\ee}{\end{equation}}

\begin{document}
\def\theequation{\arabic{section}.\arabic{equation}} 
%\preprint{}

\title{Thermal stability of stealth and de Sitter spacetimes in 
scalar-tensor gravity}

\author{Serena Giardino}
\email[]{serena.giardino@aei.mpg.de}
%%\homepage[]{Your web page}
%\thanks{}
\affiliation{Max Planck Institute for Gravitational Physics (Albert 
Einstein Institute), Callinstra{\ss}e 38, 30167 Hannover, Germany}
\affiliation{Institute for Theoretical Physics, Heidelberg University, Philosophenweg 16, 69120 Heidelberg, Germany}

\author{Andrea Giusti}
\email[]{agiusti@phys.ethz.ch}
%%\homepage[]{Your web page}
%\thanks{}
\affiliation{Institute for Theoretical Physics, ETH Zurich, 
Wolfgang-Pauli-Strasse 27, 8093, Zurich, Switzerland}

\author{Valerio Faraoni}
\email[]{vfaraoni@ubishops.ca}
%\thanks{}
\affiliation{Department of Physics \& Astronomy, Bishop's University, 
2600 College Street, Sherbrooke, Qu\'ebec, Canada J1M~1Z7}

%\date{\today}

\begin{abstract}

Stealth solutions of scalar-tensor gravity and less-known de Sitter spaces 
that generalize them are analyzed regarding their possible role as thermal 
equilibria at non-zero temperature in the new first-order thermodynamics 
of scalar-tensor gravity.  No stable equilibria are found, further 
validating the special role of general relativity as an
equilibrium state in the landscape of gravity theories, seen through the 
lens of first-order thermodynamics.

\end{abstract}

%\keywords{}

\maketitle

\section{Introduction}
\label{sec:1}
\setcounter{equation}{0}

A surprising and intriguing relationship appears to exist between 
thermodynamics and gravitation. Two seminal works showed that both 
the Einstein equations of General Relativity (GR) and the field equations of metric $f(R)$ gravity can be recovered 
from purely 
thermodynamical considerations, starting with a few assumptions 
\cite{Jacobson:1995ab,Eling:2006aw}. However, dealing with a modified 
theory of gravity requires a generalization to a non-equilibrium 
thermodynamical setting. These works put forward the idea that, in the 
landscape of gravity theories, GR could be an 
equilibrium state and 
modified gravity a non-equilibrium one. This idea was made more 
concrete by  
the recent proposal \cite{Faraoni:2021lfc,Faraoni:2021jri} of a 
first-order thermodynamics of scalar-tensor theories, with minimal 
assumptions and in a context completely different from that of spacetime 
thermodynamics \cite{Jacobson:1995ab,Eling:2006aw}. Scalar-tensor theories 
represent prototypical candidates of modified gravity and 
were 
first introduced by Brans and Dicke in \cite{BransDicke} and then extended in 
\cite{ST1, ST2, ST3}. The first-order thermodynamical proposal relies on 
interpreting the scalar contributions as an imperfect fluid 
\cite{Pimentel:1989bm,Faraoni:2018qdr,Nucamendi:2019uen} and applying a 
non-equilibrium thermodynamical description \cite{Eckart40} to it. This 
idea unexpectedly allows one to introduce a concept of ``temperature of 
gravity'' (which is clearly no physical temperature, but simply a 
temperature relative to the GR equilibrium state) and an understanding of 
the dissipative process leading gravity towards (or away from) the GR state of equilibrium. This 
proposition has been applied and tested on both different classes of 
theories (such as Horndeski gravity) and specific solutions of 
scalar-tensor theories, such as those in 
Friedmann-Lema\^itre-Robertson-Walker (FLRW) spacetime 
\cite{Giusti:2021sku,Giardino:2022sdv}.

A characteristic feature of scalar-tensor gravity is the existence of 
stealth solutions, namely solutions with the same geometry of GR solutions 
but with a nontrivial scalar field profile that does not 
contribute to the effective stress-energy tensor. Current motivation to 
study stealth solutions comes from the possibility of detecting black hole 
hair in stealth black holes through gravitational wave observations 
\cite{Takahashi:2020hso}. Indeed, ``first-generation'' scalar-tensor and 
Horndeski theories allow for stealth solutions that violate some 
assumptions of the no-hair theorems and for which the scalar field does 
not gravitate. This would in principle make it possible to observationally 
distinguish GR from scalar-tensor theories. Such solutions include stealth 
Schwarzschild (-de Sitter) black holes with a scalar field linearly 
dependent on time in the context of Horndeski and beyond-Horndeski gravity \cite{Babichev:2013cya, Babichev:2016kdt, 
Babichev:2017lmw, Kobayashi:2014eva, Motohashi:2019sen, 
BenAchour:2018dap}.

Here we are interested in stealth solutions in the 
framework of scalar-tensor thermodynamics, where they would correspond to 
different ``states of gravity'' away from the GR equilibrium, as explained in the following. Studying 
these solutions would therefore help to clarify the existence of 
equilibrium states different from GR and establish which gravity theories 
or specific solutions could approach them, extending the study of 
scalar-tensor thermodynamics to uncharted territory. 
Assessing the stability of such states is crucial: it is reasonable to 
expect that, due to the special status of the GR equilibrium state in the 
landscape of gravity theories, these other equilibria would be 
unstable, thus less relevant than GR.

The stability of certain stealth geometries has been previously studied 
with the Bardeen-Ellis-Bruni-Hwang \cite{Bardeen:1980kt, 
Ellis:1989jt,Ellis:1989ju, Ellis:1990gi, Hwang:1990am} approach for 
cosmological 
perturbations in modified gravity \cite{Hwang:1990re, Hwang:1990jh, 
Hwang:1995bv, Hwang:1996bc, 
Hwang:1996xh,Noh:2001ia}. Here, we propose 
a complementary criterion based solely on our 
thermodynamical formalism. Insights coming from thermodynamics provide 
essential guidance to both approaches, as for the stealth spacetime 
studied in \cite{Faraoni:2022jyd} with the gauge-invariant formalism. In that case, stability was assessed with the 
gauge-invariant criterion, while in the present work we mostly use the thermal criterion.

Stealth solutions include those of Refs.~\cite{Ayon-Beato:2004nzi, 
Ayon-Beato:2005yoq, 
Robinson:2006ib, Ayon-Beato:2015qfa, Alvarez:2016qky, Smolic:2017bic, 
Barjasic:2017oka, Minamitsuji:2018vuw, deRham:2019gha, Bernardo:2019yxp, 
Franzin:2021yvf, Bernardo:2020ehy, Takahashi:2020hso, Gorji:2020bfl}.  
Often these are degenerate cases of de Sitter spaces with non-constant 
scalar fields, which are not as well-known as stealth solutions 
of the field equations. de Sitter spaces with constant scalar fields are 
fixed points of the dynamical system of scalar-tensor cosmology 
\cite{Gunzig:2000kk} and are also common in GR cosmology sourced by scalar 
fields. On the contrary, de Sitter spaces with a non-constant scalar field 
are a signature of modified gravity. Both stealth solutions and de Sitter 
universes with non-constant scalar fields seem peculiar and deserve 
investigation in the thermodynamics of scalar-tensor 
gravity.

We follow the notation of Ref.~\cite{Waldbook}. 
The (Jordan frame) scalar-tensor action reads
\begin{eqnarray}
S_{\rm ST} &=& \frac{1}{16\pi} \int d^4x \sqrt{-g} \left[ \phi R 
-\frac{\omega(\phi )}{\phi} 
\, \nabla^c\phi \nabla_c\phi -V(\phi) \right] \nonumber\\
&&\nonumber\\
&\, & +S^\mathrm{(m)} \,, \label{STaction}
\end{eqnarray}
where $R$ is the Ricci scalar, the Brans-Dicke 
scalar $\phi>0$ is approximately the inverse of the effective gravitational 
coupling $G_\mathrm{eff}$, $\omega(\phi)$ is the ``Brans-Dicke coupling'', 
$V(\phi)$ is the scalar field potential, and $S^\mathrm{(m)}=\int d^4x 
\sqrt{-g} \, 
{\cal L}^\mathrm{(m)} $ is the matter action. The field equations are 
\cite{BransDicke,ST1,ST2,ST3} 
\begin{eqnarray}
R_{ab} - \frac{1}{2}\, g_{ab} R &=& \frac{8\pi}{\phi} \,  
T_{ab}^\mathrm{(m)} 
\nonumber\\
&&\nonumber\\
&\, & + \frac{\omega}{\phi^2} \left( \nabla_a \phi 
\nabla_b \phi -\frac{1}{2} \, g_{ab} 
\nabla_c \phi \nabla^c \phi \right) \nonumber\\
&&\nonumber\\
&\, &  +\frac{1}{\phi} \left( \nabla_a \nabla_b \phi 
- g_{ab} \Box \phi \right) 
-\frac{V}{2\phi}\, 
g_{ab} \,,\nonumber\\
&& \label{BDfe1}\\
(2\omega+3) \, \Box \phi &=&
\left(8\pi T^\mathrm{(m)}+ \phi \, V_{,\phi}
-2V -\omega_{,\phi} \nabla^c \phi \nabla_c \phi \right) \nonumber\\ 
&& \label{BDfe2}
\end{eqnarray}
where $R_{ab}$ is the Ricci tensor, $ T^\mathrm{(m)} 
\equiv g^{ab}T_{ab}^\mathrm{(m)} $ is the trace of the matter 
stress-energy 
tensor $T_{ab}^\mathrm{(m)}$, $\omega_{,\phi} \equiv d\omega/d\phi $ and 
$ V_{,\phi} \equiv dV/d\phi$.

\section{Thermal stability criterion}
\label{sec:2}
\setcounter{equation}{0}

Assuming $\nabla^a \phi$ to be timelike and future-oriented, it is 
used to define the four-velocity of an effective irrotational fluid 
\be
u^a = \frac{ \nabla^a \phi}{ \sqrt{ -\nabla^c \phi \nabla_c\phi}} \,.
\ee 
The effective stress-energy tensor of $\phi$ in the effective Einstein 
equations~(\ref{BDfe1}) has the form of  a dissipative fluid which, 
surprisingly, obeys Eckart's constitutive relations \cite{Eckart40} and 
leads to identifying a ``temperature of gravity'' ${\cal T}$ by
\be
{\cal KT} = \frac{ \sqrt{ -\nabla^c\phi \nabla_c\phi}}{8\pi \phi} \,,
\label{quella}
\ee
where  ${\cal K}$ is an effective thermal conductivity.  The equation 
illustrating the approach to (or the departure from) the GR 
equilibrium in scalar-tensor gravity, for theories described by the 
action~\eqref{STaction} is \cite{Faraoni:2021lfc, 
Faraoni:2021jri,Giusti:2021sku}
\be
\frac{d\left( {\cal KT}\right)}{d\tau} = 8\pi \left( {\cal KT}\right)^2 
-\Theta \, {\cal KT} +\frac{\Box\phi}{8\pi \phi} \,,\label{questa}
\ee
where $\dfrac{d}{d\tau} \equiv 
u^c\nabla_c$.

Equation~\eqref{questa} has two fixed points, 
${\cal KT}=0$ and ${\cal KT}=\rm const.>0 $. We 
explore both because, if stable, they could 
correspond to equilibrium states other than GR. 
Gravitational  theories with non-dynamical scalar fields have been shown 
to  recover 
${\cal KT}=0$ in 
\cite{Faraoni:2022doe}, while a state with ${\cal KT}=\rm const.$ that 
never approaches the GR equilibrium state was found in 
\cite{Faraoni:2022jyd} to be metastable. Here, we complement this analysis 
by studying more stealth solutions and assessing their stability 
with a new, purely thermodynamical, criterion found as 
follows.

Rewriting Eq.~(\ref{questa}) as
\be
\label{effkg}
\Box \phi-m_{\rm eff}^2 \phi=0,
\ee
where
\be
m_{\rm eff}^2 \equiv 8\pi \left[ \frac{d\left( {\cal KT}\right)}{d\tau} - 
8\pi 
\left( {\cal KT}\right)^2 + \Theta \, {\cal KT} \right] \,, 
\label{criterion}
\ee 
we 
have {\em instability} if the square of the effective mass (that we
call ``thermal mass'') is $ m_{\rm eff}^2<0 $ and {\em stability} if 
$ m_{\rm eff}^2 \geq 0$. Since ${\cal KT}$ is a scalar, this stability 
criterion is 
covariant and gauge-invariant.
This effective mass of scalar-tensor gravity differs from those explored 
in \cite{Faraoni:2009km,Helbig:1991pk}. 

The thermal stability criterion $m_{\rm eff}^2\geq 0$ is not particularly 
useful in the general thermodynamics of scalar-tensor gravity because one 
does not {\em a priori} know the quantities appearing in (\ref{criterion}). However, if one wants to assess the 
stability {\em of specific solutions} (or classes of solutions) of the 
field equations, (\ref{criterion}) is 
indeed 
suitable. This is the goal of the rest of this work. The criterion was 
used in \cite{Faraoni:2022doe} to study 
Nordstr\"om 
gravity, finding it unstable.

\section{Stealth solutions}
\label{sec:3}
\setcounter{equation}{0}

The stealth solutions we are interested in here are 
special cases where Minkowski space results not from the absence of 
matter, but from a tuned balance between matter and the 
Brans-Dicke scalar 
or, \textit{in vacuo}, between different terms in the scalar contribution 
to the stress-energy tensor. Stealth solutions like those studied in 
\cite{Ayon-Beato:2004nzi, Ayon-Beato:2005yoq, Robinson:2006ib, 
Ayon-Beato:2015qfa} are interesting since they show that Minkowski space 
is not necessarily devoid of matter, and the effect of gravitational 
coupling persists in the energy-momentum tensor even when this coupling is 
switched off.

Stealth solutions commonly encountered in the literature in the context of 
the scalar-tensor theory~(\ref{STaction}) are usually  
of two kinds:

\begin{enumerate} 

\item $g_{ab}=\eta_{ab} $ and $\phi=\phi_0 \, \mbox{e}^{\alpha \, t} $;

\item $g_{ab}=\eta_{ab} $ and $\phi=\phi_0 \, | t |^{\beta} $,

\end{enumerate}
\noindent where $\eta_{ab}$ is the Minkowski metric in Cartesian 
coordinates, $\phi_0, \alpha, \beta$ are constants, and $\phi_0>0$ so 
that gravity is always attractive.

Differentiation yields
%\footnote{{\blue Since $\frac{ d
%|t|}{dt} = \frac{|t|}{t}$, one has $\frac{  d \left( \phi_0 |t|^{\beta} 
%\right) }{dt} = \phi_0 \beta |t|^{\beta-1} \, \frac{|t|}{t} = \beta 
%\phi_0 \, \frac{|t|^{\beta}}{t} = \phi \, \frac{\beta}{t}$.}}
\be
\dot{\phi} = \phi \times 
\left\{ 
\begin{aligned}
& \alpha \, , \\
& \frac{\beta}{t}   \,,\; \; \mbox{if} \;  t\neq 0 \,,
\end{aligned}
\right.
\ee
thus the requirement of future-directed scalar gradient translates 
into the conditions
\be 
\phi > 0 \quad \mbox{and} \quad  g_{ab} \nabla^a \phi \, (\partial _t)^b < 
0 
\ee
or, for the specific scenarios above, 
\begin{eqnarray}
\label{phidot}
0 > g_{ab} \nabla^a \phi \, (\partial _t)^b &=&
 g_{ab} \left( g^{a0} 
\dot{\phi} \right) \, {\delta^b}_0 
=  g_{00} \, g^{00} \dot{\phi} = \dot{\phi} \nonumber\\
&&\nonumber\\
&=&  \phi \times 
\left\{ 
\begin{aligned}
& \alpha \, , \\
& \frac{\beta}{t} \,\,  \; \mbox{if} \; t \neq 0 \, .
\end{aligned}
\right.
\end{eqnarray}
Thus, enforcing the future orientation of the scalar field
gradient, we shall restrict to cases that satisfy the conditions

\begin{enumerate} 

\item $\alpha < 0$;

\item $\beta<0 $~ if $t>0$ or ~$\beta>0$~ if $ t<0$.

\end{enumerate}

In the first case
\be
{\cal KT} = \frac{ \sqrt{-\nabla^c \phi \nabla_c\phi}}{8\pi \phi} = 
\frac{|\alpha|}{8\pi} =\mbox{const.}>0 \,, 
\label{stealth-exp}
\ee
which means that this solution never approaches the GR equilibrium state.
If we now consider its stability from 
the point of view of first-order thermodynamics, we see that the 
effective mass is constant and given by  
\begin{eqnarray}
m_{\rm eff}^2 &=& \frac{\Box \phi}{\phi} = 
\frac{\partial^{\mu}\partial_{\mu}\phi}{\phi} \nonumber\\
&&\nonumber\\
&=&  \frac{\partial^{\mu} \left(  \alpha \, 
{\delta^0}_{\mu} \phi \right)}{\phi} = -\alpha^2 <0 \,,
\end{eqnarray}
which makes this stealth solution {\em unstable}.  A stealth 
solution of this 
type was assessed in \cite{Faraoni:2022jyd} with 
the  gauge-invariant criterion 
for cosmological perturbations and shown to be a metastable state. \\

In the second case, $\beta=1$ and $\beta=2$ are the most relevant 
situations encountered in the literature. Therefore, according to our 
conventions, in order to have $G_\mathrm{eff} = \phi^{-1}>0$ and 
$u^a = \nabla^a \phi / \sqrt{-\nabla^c\phi \nabla_c\phi}$ future-oriented, it must be
$\phi_0>0$ in conjunction with $t<0$ if $\beta > 0$.

Then, if $\beta > 0$ the effective gravitational coupling behaves as
\be
G_\mathrm{eff}=\frac{1}{\phi} = \frac{1}{\phi_0 \, |t|^\beta} \to +\infty \quad 
\mbox{as} \; t\to 0^{-} \, ,
\ee
the effective temperature of gravity (\ref{quella}) is 
\be
{\cal KT} = \frac{ 
\beta}{8\pi |t|} \to 
+\infty \quad 
\mbox{as} \; t\to 0^{-} \, ,
\ee
and the effective mass reads
\be
m_\mathrm{eff}^2 =  \frac{\Box\phi}{\phi} = - \frac{\beta (\beta - 1)}{t^2}  \, .
\ee

If $\beta=1$, we get $m_{\rm eff}^2 = 0$. Therefore this constant ``mass'' solution is {\em marginally stable}.
As $t\to 
0^{-}$, we approach a singularity of the theory where $G_\mathrm{eff}\to + 
\infty$, ${\cal KT} \to +\infty$, gravity becomes infinitely strong and 
deviates from GR drastically. Indeed, nothing could be further from a GR 
situation than infinitely strong gravity with Minkowski spacetime! This 
solution matches the idea that singularities are ``hot'' in the sense of 
the thermodynamics of scalar-tensor gravity 
\cite{Faraoni:2021lfc,Faraoni:2021jri}. This situation is stable 
according to the 
thermal stability criterion~(\ref{criterion}). Hence, barring instabilities 
of a different nature, one expects this behaviour to occur in nature if 
singularities are present. The implication is that the GR equilibrium state is 
not always approached and gravity  
indeed departs from GR near singularities. Of course, the final theory of 
gravity should remove singularities, but it is  clear that scalar-tensor 
gravity is not this 
final theory since it does contain spacetime singularities and 
singularities of $G_\mathrm{eff}$.

The situation where $\beta=2$, exemplified in Sec.~\ref{nariai}, entails
$m_\mathrm{eff}^2 =  - 2/t^2 <0$, meaning \textit{instability} from the 
thermal point of view, while ${\cal KT} = 1/4\pi |t|$ and $G_\mathrm{eff} = 1/\phi_0 t^2$ both diverge as $t\to 0^{-}$, thus departing from GR at the singularity of $G_\mathrm{eff}$. In our formalism the $t>0$ branch of the solution is not meaningful.

Most exact solutions of Brans-Dicke theories in cosmology 
exhibit the 
power-law behaviour $\phi=\phi_0\,t^{\beta}$ \cite{Faraoni:2017ecj}, such 
as those found by O'Hanlon and Tupper \cite{OHanlon:1972ysn} and Nariai 
\cite{Nariai,Johri:1994rw}. These were studied from the point of view of 
first-order thermodynamics in \cite{Giardino:2022sdv}, and in \ref{oht} 
and \ref{nariai} we consider two degenerate cases of such solutions that 
reduce to a Minkowski background with a non-trivial scalar field profile.

Other types of stealth solutions with Minkowski metric and non-trivial 
scalar include those found for a nonminimally coupled $\phi$ 
\cite{Ayon-Beato:2005yoq}, where the field is inhomogeneous, wave-like, 
and 
does not gravitate. Their stability was studied in \cite{Faraoni:2010mj} 
using the Bardeen-Ellis-Bruni-Hwang gauge-invariant formalism for 
cosmological perturbations \cite{Bardeen:1980kt, 
Ellis:1989jt,Ellis:1989ju, Ellis:1990gi, Hwang:1990am}, showing mixed 
stability results depending on the specific choice of parameters.
These solutions either do not correspond to future-oriented 
four-velocity $u^c$, or are very cumbersome to discuss because 
$\nabla^a\phi$ is timelike only in very restricted spacetime 
regions and for special combinations of their parameters. Therefore, they will not be examined here.

\subsection{O'Hanlon \& Tupper (OHT) solution with $\omega=0$}
\label{oht}

The O'Hanlon \& Tupper spatially flat FLRW solution of Brans-Dicke cosmology is 
obtained from the action~(\ref{STaction}) for $\omega=\mbox{const.}>-3/2 
$ and $\omega \neq -4/3$ and  $V=0$ \cite{OHanlon:1972ysn}. The scale 
factor and scalar field read
\begin{eqnarray}
a(t) &=& a_0 \left( \frac{t}{t_0} \right)^{q_{\pm}} \,,\label{OHT-1}\\
&&\nonumber\\
\phi(t) &=& \phi_0 \left( \frac{t}{t_0} \right)^{s_{\pm}}  \,,\label{OHT-2}
\end{eqnarray}
with 
\begin{eqnarray}
q_{\pm} &=& \frac{\omega}{3(\omega+1) \mp \sqrt{ 3(2\omega+3)}} \,, \label{OHT-3}\\
&&\nonumber\\
s_{\pm} &=& \frac{ 1 \mp \sqrt{3(2\omega+3)}}{3\omega+4}  \,,\label{OHT-4}
\end{eqnarray}
and $ 3q_{\pm}+s_{\pm}=1 $. 
This solution has a ``hot'' 
singularity at $t\rightarrow 0^+ $, where Brans-Dicke theory departs from the GR 
behaviour. Although the value $\omega=0$ was not 
contemplated in 
\cite{OHanlon:1972ysn}, it is straightforward to check that it corresponds 
to a Minkowski space solution of the equations of vacuum Brans-Dicke 
cosmology with $V=0$, $q=0$, $a(t)=1$, and linear scalar field $\phi(t) 
=\phi_0 \, t$ (choosing $t_0=1$ for 
convenience). This is a {\em bona fide} stealth solution, which 
could 
have been introduced in Ref.~\cite{OHanlon:1972ysn} long before solutions 
with this name were noticed and appreciated 
\cite{Ayon-Beato:2004nzi, Ayon-Beato:2005yoq, Robinson:2006ib, Ayon-Beato:2015qfa, 
Alvarez:2016qky, Smolic:2017bic, Barjasic:2017oka, Minamitsuji:2018vuw, deRham:2019gha, 
Bernardo:2019yxp, Franzin:2021yvf, Bernardo:2020ehy, Takahashi:2020hso, Gorji:2020bfl}. 
In order for the four-velocity to be future-oriented and 
for $G_\mathrm{eff}$ to be positive, it must be $\phi_0<0$ and $t<0$. This situation is akin to case 2. with $\beta = 1$ considered 
above, hence the $\omega=0$ O'Hanlon \& Tupper solution turns out to be \textit{marginally stable} according to the thermal stability criterion.\footnote{In the analysis at the beginning of Sec.~\ref{sec:3}, we conventionally denoted $\phi (t) =\phi_0 \, |t|^\beta$ with $\phi_0>0$. In this section we instead employ the usual notation that can be found in the literature, {\em i.e.}, $\phi (t) =\phi_0 \, t^\beta$, where $\phi_0$ and $t$ can both be either positive or negative, provided that $\phi$ remains positive. \label{footnote}} 
This universe has 
\be
{\cal KT}=\frac{1}{8\pi |t|}   \to  +\infty
\ee
as $t\to 0^{-}$, deviating from GR.\\

\subsection{Nariai solution with $\omega=-1/2$}
\label{nariai}

The Nariai solution \cite{Nariai, Johri:1994rw} is a particular power-law 
solution for a 
$K=0$ FLRW universe with perfect fluid matter that has $P=\left( 
\gamma-1 \right) \rho$ (with $\gamma=$~const.), $V(\phi)=0$ and $\omega\neq 
-4 \left[ 3\gamma \left(2-\gamma \right) \right]^{-1}<0$. Here we are 
interested in a cosmological constant fluid with $\gamma=0$, 
$P^\mathrm{(m)}=-\rho^\mathrm{(m)}$, and
\begin{eqnarray}
    &&a(t)=a_0 \left( 1+\delta t \right)^{\omega+1/2} \,,\\
    &&\nonumber\\
    &&\phi(t)=\phi_0 \left( 1+\delta t \right)^2 \,,\\
    &&\nonumber\\
    &&\delta=\left[ \frac{32\pi\rho_0}{\phi_0} 
\frac{1}{\left( 6\omega+5 \right) \left( 2\omega+3 \right)}\right]^{1/2} \,.
\end{eqnarray}

This solution is an attractor in phase space and was used in the extended 
inflationary scenario \cite{La:1989za, La:1989st}. For $\omega=-1/2$, $ \delta =\sqrt{ 8\pi \rho_0/\phi_0}$, the 
scale factor is constant and $H=0$, making this a Minkowski 
stealth solution with non-trivial (polynomial) scalar field profile. It is a straightforward generalisation of the type 2. stealth solutions described above. \footnote{Here again we implicitly adapted our notation to the one which is typically employed in the literature. See footnote \ref{footnote}.}
It must be $\phi_0>0, \left( 1+\delta t \right)<0$ and 
\be
{\cal K}{\cal T} = \frac{\delta}{4\pi |1+\delta t|} \to +\infty
\ee
as $\left( 1+\delta t\right) \to 0^{-}$. In the far past $t \to - \infty$, 
${\cal 
K}{\cal T}\to 0$ and GR is approached, but the instability prevents this 
state from being an equilibrium alternative to GR. In fact, the thermal stability criterion
yields
\be
m^2_{\rm eff}=\frac{\Box\phi}{\phi}= - \, 
\frac{2 \delta^2}{\left( 1+\delta t \right)^2}<0 
\ee
and this solution is thermally \textit{unstable}.

\section{de Sitter space solutions}
\label{sec:4}
\setcounter{equation}{0}

Other common solutions of scalar-tensor gravity are de Sitter ones with line element
\be
ds^2 =-dt^2 +a_0^2 \, \mbox{e}^{2H_0 t} \left( dx^2 +dy^2 +dz^2 \right) 
\ee
in comoving coordinates, with scale factor 
$a(t)=a_0 \, \mbox{e}^{H_0t}$, where $a_0, H_0$ are constants. 

In GR with a minimally coupled scalar field as the only matter source, the 
only possible de Sitter spaces are obtained for a constant scalar field, 
$\left( H, \phi \right)=\left( H_0, \phi_0 \right)$, with both $H_0$ and 
$\phi_0$ constant. In spatially flat FLRW cosmology, the independent 
dynamical variables are\footnote{In the field equations for spatially flat 
FLRW universes, the scale factor only appears in the 
combination $H \equiv \dot{a}/a$.} 
$\left( H, \phi \right) $ and the phase space is a 2-dimensional subset of 
the 3-dimensional space $ \left( H, \phi, \dot{\phi} \right)$ 
identified 
by the Hamiltonian constraint. This 2-dimensional subset is analogous to 
an energy surface in point particle mechanics 
\cite{Faraoni:2005vc,deSouza:2007zpn}.  The 
points $\left( H_0, \phi_0 \right)$ are then all the equilibrium points of 
the dynamical system.

For spatially flat FLRW universes in scalar-tensor cosmology, the 
independent variables are still $H$ and $\phi$ and there can be fixed points $\left( H_0 , \phi_0 \right)$ of this dynamical system. The 
structure of the phase space and the fixed points for specific 
scalar-tensor theories are discussed extensively in \cite{Faraoni:2005vc} 
and \cite{Bxl, Gunzig:2000kk}, respectively. Gauge-invariant criteria for 
the stability of these de Sitter fixed points (and of their degenerate 
Minkowski cases) are given in \cite{Faraoni:2004dn, 
Faraoni:2005ie, Faraoni:2005vk, Faraoni:2006ik,Faraoni:2007yn}. In 
addition to de Sitter fixed points, in scalar-tensor cosmology there can be de Sitter spaces with non-constant scalar 
field, usually exponential or power-law in time. Since these are only 
admissible in modified gravity and not in GR, they are interesting for first-order 
thermodynamics. Degenerate cases of such de Sitter solutions can reproduce 
Minkowski space with a non-trivial scalar field and are therefore another 
kind of stealth solutions similar to those of
the previous section.

\subsection{de Sitter solutions of scalar-tensor gravity}

This type of solution, known in many scalar-tensor theories, is found 
starting from the action~\eqref{STaction} and reads
\begin{eqnarray}
H &=& H_0 = \mbox{const.} \,,\\
&&\nonumber\\
\phi(t) &=& \phi_0 \, \mbox{e}^{\alpha \, t} \,,
\end{eqnarray}
with $\phi_0$ a positive constant. The constants $H_0$ and 
$\alpha$ are related to 
the parameters of the specific scalar-tensor theory. Although these 
solutions have been known  for a long time, here we consider them from the 
novel point of view of scalar-tensor 
thermodynamics. 

In order to get a future-directed four-velocity of the effective $\phi$-fluid and an attractive gravitational interaction we need to require, again, that
\be 
\phi > 0 \quad \mbox{and} \quad  g_{ab} \nabla^a \phi \, (\partial _t)^b < 
0 \, , 
\ee
which implies $\phi_0 > 0$ and $\alpha < 0$.

We have (as 
in~(\ref{stealth-exp})) 
\be
{\cal KT}= \frac{|\alpha|}{8\pi}= \mbox{const.}
\ee
and this solution remains away from the zero-temperature GR 
state of equilibrium at all times. Is it thermally stable? We find
\begin{eqnarray}
\label{expcrit}
m_{\rm eff}^2 &=& \frac{\Box\phi}{\phi} = \frac{ -\left( \ddot{\phi}+3H_0 
\dot{\phi}  \right)}{\phi} = -\alpha\left( \alpha+3H_0 \right) \nonumber\\
&&\nonumber\\
&=& |\alpha|\left( 3H_0-|\alpha| \right) \,;
\end{eqnarray}\\
therefore,  we have \textit{stability} for $3H_0 \geq |\alpha|$ and 
\textit{instability} for $|\alpha| > 3H_0$. 

In particular, it is clear that exponentially contracting FLRW universes 
($H_0<0$) are always unstable. This conclusion, obtained with simple 
considerations in scalar-tensor thermodynamics, 
matches the result found in the literature on scalar-tensor cosmology 
\cite{Faraoni:2004dn} with a dynamical systems analysis which requires the 
complete specification of the theory.

\subsubsection{Kolitch solutions of vacuum Brans-Dicke 
cosmology with cosmological constant}

Kolitch \cite{Kolitch:1994kr} found solutions of vacuum Brans-Dicke 
cosmology with positive cosmological constant $\Lambda$, equivalent to the 
linear potential $V(\phi) = 2\Lambda \phi $. These solutions were 
previously noted in \cite{Barrow:1990nv,RomeroBarros93} and read
\begin{eqnarray}
a(t) &=& a_0 \exp \left[ \pm \left( \omega+1 \right) \sqrt{ 
\frac{2\Lambda}{(2\omega+3)(3\omega+4)}} \, t \right] \,,\nonumber\\ 
&&\label{Kolitch1}\\
\phi(t) &=& \phi_0 \exp \left[ \pm \sqrt{ 
\frac{2\Lambda}{(2\omega+3)(3\omega+4)}} \, t \right] \,.\label{Kolitch2}
\end{eqnarray}
For $\omega=-1$, they reduce to the stealth solution with 
\be
H=0 \,, \quad \quad a(t)=1 \,, \quad\quad \phi(t)=\phi_0 \, 
\mbox{e}^{ 
\pm 
\sqrt{2\Lambda} \, t, } 
\ee
where, again, we must choose the lower sign to have a future-oriented 
four-velocity. This solution deviates from GR at 
all times since 
${\cal KT}=\mbox{const.}>0$, but it corresponds to $m_{\rm 
eff}^2 =-\alpha^2 <0$ and is
\textit{unstable}. Its stability has also 
been studied with respect to both homogeneous and inhomogenous metric 
perturbations in 
\cite{Faraoni:2010mj}, where the solution with the upper 
sign is found to be stable and the one with the lower sign unstable. However, the solution with the upper sign cannot be analysed in the framework of scalar-tensor thermodynamics since it entails a past-oriented $\nabla^a \phi$.

Let us consider now the de Sitter spaces~(\ref{Kolitch1}), 
(\ref{Kolitch2}) for $\omega\neq -1$: taking the lower sign we have 
\be
H_0 = -\left( \omega+1\right) \sqrt{ 
\frac{2\Lambda}{(2\omega+3)(3\omega+4)}} \equiv 
- \left( \omega+1\right) {\rm C}
\ee
and
\be
\alpha=- \sqrt{\frac{2\Lambda}{(2\omega+3)(3\omega+4)}} \equiv - {\rm C} \, ,
\ee
where ${\rm C}$ is a positive real constant if $\omega < -3/2$ and $\omega > -4/3$. Therefore, the effective mass reads
\begin{eqnarray}
m_{\rm eff}^2&=& |\alpha|\left( 3H_0-|\alpha| \right) = -{\rm C}^2 \left( 3 \omega + 4 \right)
\end{eqnarray}
Then, if $\omega < -3/2$ we have an expanding de Sitter universe which is thermodynamically \textit{stable}, although the scalar field for such values of the coupling is phantom and therefore suffers from different types of instabilities \cite{Faraoni:2004pi}. Other configurations are otherwise \textit{unstable}.

\subsubsection{O'Hanlon \& Tupper solution in the $\omega\rightarrow 
-4/3 $ limit}

It is often mentioned in the literature that the O'Hanlon \& Tupper 
solution (\ref{OHT-1})-(\ref{OHT-4}) approaches de Sitter space in the 
limit $\omega\rightarrow -4/3$, recovering
\begin{eqnarray}
\label{oht43}
a(t) &=& a_0 \exp{(H_0\,t)} \,,\\ 
\nonumber
&&\\
\label{oht432}
\phi(t) &=& \phi_0\exp{(-3H_0\,t)} \,,
\end{eqnarray}
with $H_0$ a positive constant. Technically, this statement is not 
accurate since 
the above result is recovered by simultaneously choosing the values $q_+$ 
and $s_-$ of the exponents, which correspond to two distinct solutions. 
However, the solution above is the only de Sitter one for flat FLRW and 
vacuum \cite{Faraoni:2004pi}. Given that $\alpha<0$, the velocity of the 
scalar field fluid is future-oriented and $3H_0-|\alpha|=0$, so this 
solution is \textit{marginally stable} according to the thermal criterion.

This solution describes expanding universes for which the effective fluid four-velocity is only future-oriented. These expanding universes are unstable with respect to tensor modes, as can 
be concluded using the Bardeen-Ellis-Bruni gauge-invariant formalism for 
cosmological perturbations 
\cite{Bardeen:1980kt, Ellis:1989jt,Ellis:1989ju, Ellis:1990gi, 
Hwang:1990am} in Hwang's version adapted to modified 
gravity \cite{Hwang:1990re, Hwang:1990jh, Hwang:1995bv, Hwang:1996bc, 
Hwang:1996xh,Noh:2001ia}.  The 
relevant equations are summarized in Appendix~\ref{Appendix}. We only 
need Eq.~(\ref{Aeq:28}) for the gauge-invariant variable $H_T$ 
associated with the tensor modes which, in the background (\ref{oht43}) and 
(\ref{oht432}), becomes
\be
\ddot{H}_T +\left( 3H+ \frac{\dot{\phi}}{\phi} \right) \dot{H}_T + 
\frac{k^2}{a^2(t)} \, H_T =0 \,,
\ee
where $k$ is the mode's momentum and the coefficients are given by the 
unperturbed $a(t)$ and $\phi(t)$, which yields $ 3H+\dot{\phi}/\phi=0$ to 
zero order. With $H_0>0$, the asymptotic equation at late times $t\to 
+\infty$ reduces  to
\be
\ddot{H}_T +\frac{k^2}{a^2} \, H_T  \simeq \ddot{H}_T=0 \,,
\ee
with linear solution $H(t)= \alpha\, t+ \mbox{const.}$ The tensor 
perturbation diverges and this universe is \textit{unstable}.

\begin{widetext}

\setlength{\tabcolsep}{15pt}
\renewcommand{\arraystretch}{2.5}
\begin{table}[ht!]
\centering
 \begin{tabular}{||c| c| c||} 
 \hline
 {\bf Solution} & {\bf Type} & {\bf Thermal Stability} 
 \\ [0.5ex] 
 \hline\hline
 OHT $\omega=0$ & Minkowski stealth & marginally stable (departs 
from GR  as $t\to 0^{-} $) 
\\
\hline
 Nariai $\omega=-1/2$ & Minkowski stealth & unstable 
 \\
\hline
 Kolitch $\omega=-1$  & Minkowski stealth &  unstable \\
\hline
Kolitch $\omega<-3/2$ & de Sitter 
&  stable (but $\phi$ phantom)
 \\
\hline
 OHT $\omega\rightarrow -4/3$ & de Sitter & marginally stable 
 \\
\hline
 $f(R)$ gravity  & (Anti-)de Sitter, Minkowski 
& marginally stable
 \\
 [1ex] 
 \hline
 \end{tabular}
\caption{Summary of the analytical solutions studied and their thermal stability.}
\label{Table:1}
\end{table}
\end{widetext}

\subsection{Constant curvature spaces in $f(R)$ gravity}

Metric $f(R)$ gravity is a subclass of scalar-tensor theories described by the action
\be
S_{f(R)}=\frac{1}{16\pi} \int d^4 x \sqrt{-g} f(R) +S^\mathrm{(m)}
\ee
and is equivalent \cite{Sotiriou:2008rp, DeFelice:2010aj, Nojiri:2010wj}  
to a Brans-Dicke theory with $\phi= f'(R)$ (a prime denotes 
differentiation with respect to $R$), $\omega=0$, and the potential
\be
V(\phi) =Rf'(R)-f(R) \Bigg|_{f'(R)=\phi} \,.
\ee
Assuming that $\nabla^c R$ is timelike and future-oriented, the 
effective dissipative fluid 
associated with $f(R)$ gravity has \cite{Faraoni:2021lfc}
\be 
{\cal KT} = \frac{ f''(R) \sqrt{ -\nabla^c R \nabla_c 
R}}{8\pi f'(R) } 
\,,
\ee
where it is required that $f'(R)>0$ in order for the effective 
gravitational coupling $G_\mathrm{eff} =1/\phi$ to be positive and for  
the graviton to carry positive kinetic energy, while $f''(R) \geq 0$ is 
required for local stability \cite{Faraoni:2006sy} (here $\nabla^c 
\phi$ is timelike and future-oriented if $\nabla^c R$ is). 

The fact that the effective Brans-Dicke scalar field $\phi$ in $f(R)$ 
gravity is tied so intimately with the Ricci scalar makes all constant 
curvature spaces in these theories zero-temperature states 
indistinguishable from GR, because this means that 
$\phi=f'(R)=$~const. and $\nabla_c \phi$ vanishes identically, together 
with ${\cal KT}$. Furthermore, these states are \textit{(marginally) stable} 
in our thermal sense because 
$\Box \phi=0 $ and the effective mass $m_{\rm eff}^2= \Box\phi/\phi$ also vanishes 
identically.

The condition $m^2_\mathrm{eff} \geq 0 $ for the thermal stability of $f(R)$ 
gravity does 
not coincide with the stability condition of de Sitter space with respect 
to first order local perturbations, obtained in a gauge-invariant way 
(\cite{Faraoni:2007yn} and references therein),
\be
\left(f_0' \right)^2-2f_0 f_0'' \geq 0 \,,
\ee
where a 
zero 
subscript denotes a  quantity evaluated on the de Sitter 
background. Therefore, the thermal 
stability condition $m^2_\mathrm{eff} \geq 0$ does not 
necessarily 
coincide with other stability notions, as could be expected. Indeed, also 
in Newtonian systems and in GR one has different notions of stability 
(thermal, dynamical, {\em etc.}) and the thermodynamics of modified 
gravity evidently cannot account for all possible notions of 
stability.

\section{Conclusions}
\label{sec:5}
\setcounter{equation}{0}
In this work, we studied the states of gravity corresponding to ${\cal KT}=\rm const.$, which are fixed points of the effective heat equation describing the approach to (or departure from) equilibrium \eqref{questa}, in the context of first-order thermodynamics \cite{Faraoni:2021jri}.
These states, away from the GR equilibrium, correspond to different types of stealth solutions, 
%and we assessed their stability by making use of a new thermal criterion~(\ref{criterion}).}
which are not admitted by the Einstein equations and are thus a 
signature of alternative gravity \cite{Ayon-Beato:2004nzi, 
Ayon-Beato:2005yoq, Robinson:2006ib, Ayon-Beato:2015qfa, Alvarez:2016qky, 
Smolic:2017bic, Barjasic:2017oka, Minamitsuji:2018vuw, deRham:2019gha, 
Bernardo:2019yxp, Franzin:2021yvf, Bernardo:2020ehy, Takahashi:2020hso, 
Gorji:2020bfl}. 

Specifically, we studied the scalar 
field profiles 1. $\phi=\phi_0 \, \mbox{e}^{\alpha \, t}$ (with 
$\alpha<0$) and 2. $\phi=\phi_0 \, 
| t |^{\beta}$ (with $t>0, \beta<0$ or with $t<0, 
\beta>0$), common in the literature. The first case has ${\cal KT}= 
\mbox{const.}>0$, which would 
correspond to a state of equilibrium at positive temperature. However, 
this state is unstable according to a new, purely thermal, criterion that we find~(\ref{criterion}). This criterion does not necessarily go hand-in-hand with other 
stability criteria, which should not come as a surprise, since a physical 
system can be subject to instabilities of different nature, with different time scales. Sometimes instability in the thermal 
sense~(\ref{criterion}) is accompanied by instability with respect to 
gravitational perturbations; however, this coincidence should not 
always be 
expected. 

In any case, stable equilibrium states of gravity with ${\cal 
KT}=\rm const.$ either do not exist or are fragile and easily destroyed by 
perturbations ({\em i.e.}, metastable).

Stealth solutions with a linear scalar field profile, as in the second case, require caution because, combining the requirements that $G_\mathrm{eff}>0$ and 
that the effective $\phi$-fluid four-velocity $u^a$ be future-oriented (essential when discussing dissipation associated with an arrow 
of time), one finds a singularity of the effective gravitational coupling 
at $t=0$, which can justly be regarded as a ``thermodynamical'' 
singularity of scalar-tensor gravity. These spaces are stable according to the thermal criterion and are not destroyed by perturbations (as 
far as scalar-tensor gravity applies), but ${\cal KT}$ diverges at this 
singularity, as it does in ordinary spacetime singularities, signalling a drastic deviation from GR predicted in 
\cite{Faraoni:2021lfc, Faraoni:2021jri}. This result reinforces the idea 
that gravity strongly deviates from GR at singularities, but now the 
concept of ``thermodynamical singularity'' is extended to include also singularities of 
the effective gravitational coupling $G_\mathrm{eff}$. These 
considerations, of course, do not solve the spacetime singularity problem of 
relativistic gravity; the temperature ${\cal T}$ introduced by 
scalar-tensor thermodynamics is relative to the GR state and measures the 
distance of the actual state of gravity from the GR state of equilibrium 
at ${\cal KT}=0$, which is still affected by the spacetime singularity 
problem.

The realization that stealth solutions of scalar-tensor gravity are often 
degenerate cases of de Sitter universes with non-constant Brans-Dicke-like 
scalar field prompts the consideration of these spaces (Sec.~\ref{sec:4}). 
It is intriguing that the cosmic no-hair theorem (when valid) can be seen in a new 
light from the point of view of scalar-tensor thermodynamics. (The validity, or lack thereof, of 
cosmic no-hair in various scalar-tensor gravities will be examined from 
the thermal point of view in future work). On the one hand, de Sitter 
spaces with constant scalar field can be attractors of the cosmological 
dynamics (even starting with anisotropic Bianchi models) but, when $\phi$ 
is constant, ${\cal KT}$ vanishes and gravity 
reduces to its zero-temperature GR state of equilibrium.\footnote{Indeed, de 
Sitter spaces with constant scalar field are common attractors in GR 
according to Wald's theorem (\cite{Wald:1983ky}, see 
\cite{Goldwirth:1991rj} for a review).} On the other hand, de Sitter spaces with 
non-constant scalar field are known to occur in various scalar-tensor gravities (where they are not attractors of the cosmological dynamics) but are impossible in GR and are a signature of alternative gravity. In this sense, they can be regarded as generalizations of stealth solutions \cite{Motohashi:2019ymr} and as such they were studied here from the point of view of first-order thermodynamics.

The results obtained for the solutions of scalar-tensor gravity 
analyzed here are summarized in Table~\ref{Table:1}. Overall, the two general principles of first-order thermodynamics of 
scalar-tensor gravity are confirmed: {\em i})~gravity 
deviates wildly from GR 
near spacetime singularities and near singularities of the gravitational 
coupling; {\em ii})~the convergence of gravity to GR at late times is 
marked by ${\cal K}{\cal T} \to 0$. No states of equilibrium ${\cal 
K}{\cal T}=$~const. other than GR (corresponding to ${\cal K}{\cal T}=0$) 
have been found here, except for solutions that are unstable according to 
various criteria and are, therefore, physically irrelevant. This result reinforces the special role of general relativity as an
equilibrium state in the landscape of gravity theories, seen through the 
lens of first-order thermodynamics.
The results above will be useful in the following developments of the first-order thermodynamical formalism.

\begin{acknowledgments}

S.~G. thanks Jean-Luc Lehners at AEI Potsdam for hospitality. A.~G. is 
supported by the European Union's Horizon 2020 research and innovation 
programme under the Marie Sk\l{}odowska-Curie Actions (grant agreement 
No.~895648--CosmoDEC). The work of A.~G. has also been carried out in the 
framework of the activities of the Italian National Group of Mathematical 
Physics [Gruppo Nazionale per la Fisica Matematica (GNFM), Istituto 
Nazionale di Alta Matematica (INdAM)]. V.~F. is supported by the Natural 
Sciences \& Engineering Research Council of Canada (grant 2016-03803).

\end{acknowledgments}

\appendix\section{Gauge-invariant perturbations for scalar-tensor 
cosmology}  
\label{Appendix}

\renewcommand{\theequation}{A.\arabic{equation}}

Consider the modified gravity described by the action  
\begin{equation}
S=\int  d^{4}x\,\sqrt{-g}\,\left[\frac{f(\phi,R)}{2}
-\frac{ \bar{\omega}(\phi)}{2} \, \nabla^c \phi\nabla_c \phi
-V\left(\phi\right)\right] \label{Aeq:2.1}
\end{equation}
and a spatially flat unperturbed FLRW  universe with line element
\be \label{FLRWmetric}
ds^2=-dt^2+a^2(t) \left( dx^2+dy^2 +dz^2 \right) \,.
\ee
The unperturbed field equations are  
\begin{eqnarray}
H^{2} & = & \frac{1}{3F} \left(\frac{ \bar{\omega}}{2} \, \dot{\phi}^{2} 
+\frac{RF}{2}-\frac{f}{2}+V-3H\dot{F}\right)\,\,,
\label{Aeq:12} \\
&&\nonumber\\
\dot{H} & = & -\frac{1}{2F}\left( \bar{\omega} \, \dot{\phi}^{2} 
+\ddot{F}-H\dot{F}\right)\,\,,\label{Aeq:13}\\
&&\nonumber\\
\ddot{\phi} & + & 3H\dot{\phi}+\frac{1}{2 \bar{\omega}} 
\left(\frac{d \bar{\omega}}{d\phi} \, \dot{\phi}^{2}-\frac{\partial f}{ 
\partial\phi}+2 \, \frac{dV}{d\phi}\right)=0\,\,,\label{Aeq:14}
\end{eqnarray}
where an overdot denotes differentiation with respect to the comoving 
time $t$,  $H\equiv\dot{a}/a$ is the Hubble function, and $F\equiv\partial 
f/\partial R$. Quantities denoted with $A,\, 
B,\, H_{L}$, and $H_{T}$ define the metric  perturbations in the 
Bardeen-Ellis-Bruni-Hwang formalism 
\cite{Bardeen:1980kt, 
Ellis:1989jt,Ellis:1989ju, Ellis:1990gi, Hwang:1990am} according to 
\begin{eqnarray}
g_{00} & = & -a^{2}\left(1+2AY\right)\,\,,\label{Aeq:15}\\
&&\nonumber\\
g_{0i} & = & -a^{2}BY_{i}\,\,,\label{Aeq:16} \\
&&\nonumber\\
g_{ij} & = & a^{2}\left[h_{ij}\left(1+2H_{L}\right) 
+2H_{T}Y_{ij}\right]\,,\label{Aeq:17}
\end{eqnarray}
where $h_{ij}$ is the 3-metric of the unperturbed FLRW space seen by the 
comoving observer,  
the scalar harmonics $Y$ satisfy the eigenvalue problem 
$\bar{\nabla}_{i}\bar{\nabla}^i Y=-k^{2}Y$ with   
eigenvalue $k$,  and $\bar{\nabla_{i}}$ is the covariant derivative operator 
of  $h_{ij}$.  The  vector and tensor  harmonics $Y_{i}$ and $Y_{ij}$ 
satisfy
\begin{eqnarray}
Y_{i} &=& -\frac{1}{k} \, \bar{\nabla}_{i}Y\,\,,\label{Aeq:18}\\
&&\nonumber\\
Y_{ij} &=& \frac{1}{k^2} \, \bar{\nabla}_{i} 
\bar{\nabla}_{j}Y+\frac{1}{3}Yh_{ij}\,\,.\label{Aeq:19}
\end{eqnarray}
\begin{eqnarray}
\Phi_{H} & = & H_{L}+\frac{H_{T}}{3} +\frac{\dot{a}}{k} \left(B 
-\frac{a}{k}\,\dot{H}_{T}\right)\,,\label{Aeq:20}\\
&&\nonumber\\
\Phi_{A} & = & A+\frac{\dot{a}}{k}\left(B 
-\frac{a}{k}\,\dot{H}_{T}\right) 
+\frac{a}{k}\left[\dot{B}-\frac{1}{k} 
\left(a\dot{H}_{T}\right)\dot{} \,\right]\,,\nonumber\\
&&  \label{Aeq:21}
\end{eqnarray}
are the Bardeen gauge-invariant potentials 
\cite{Bardeen:1980kt},
\begin{equation}
\Delta \phi=\delta\phi+\frac{a}{k} \, \dot{\phi}\left(B 
-\frac{a}{k} \, \dot{H}_{T}\right) \label{Aeq:22}
\end{equation}
is the Ellis-Bruni variable \cite{Ellis:1989jt,Ellis:1989ju},  and similar 
relations 
define the other gauge-invariant variables  $\Delta f,\,\Delta F,$ and 
$\Delta R$. 
We refer the reader to Refs.~\cite{Hwang:1990re, Hwang:1990jh, 
Hwang:1995bv, Hwang:1996bc, 
Hwang:1996xh,Noh:2001ia} for the 
complete set of equations for the 
gauge-invariant perturbations. Here we only need the equation for the 
tensor modes
\be \label{Aeq:28} 
\ddot{H}_T +\left( 3H+ \frac{\dot{F}}{F} 
\right) 
\dot{H}_T +\frac{k^2}{a^2} \, H_T=0 \,,
\ee
which is used in Sec.~\ref{sec:4}.

%\bibliographystyle{apsrev4-2}
%\bibliography{ref}{}
%\begin{thebibliography}{99}%
%

\end{document}